\documentclass[10pt,superscriptaddress,twocolumn,amsmath,amssymb,aps,prl,showpacs]{revtex4-1}
\usepackage{mathrsfs}
\usepackage{graphicx}
\usepackage{dcolumn}
\usepackage{bm}

\usepackage{amssymb}
\usepackage{amsmath}

\newcommand{\be}{\begin{equation}}
\newcommand{\ee}{\end{equation}}
\newcommand{\bea}{\begin{eqnarray}}
\newcommand{\eea}{\end{eqnarray}}

\begin{document}
\title{Universal Relations for a Fermi Gas Close to a $p$-wave Interaction Resonance}
\author{Zhenhua Yu}
\affiliation{Institute for Advanced Study, Tsinghua University,
Beijing, 100084, China}

\author{Joseph H. Thywissen}
\affiliation{Department of Physics, University of Toronto, M5S 1A7 Canada}

\author{Shizhong Zhang}
\affiliation{Department of Physics and Center of Theoretical and
Computational Physics, The University of Hong Kong, Hong Kong,
China}

\date{\today}
\begin{abstract}
We investigate the properties of a spinless Fermi gas close to a $p$-wave interaction resonance. We show that the effects of interaction near a $p$-wave resonance are captured by two contacts, which are related to the variation of energy with the $p$-wave scattering volume $v$ and with the effective range $R$ in two adiabatic theorems. Exact pressure and virial relations are derived. We show how the two contacts determine the leading and sub-leading asymptotic behavior of the momentum distribution ($\sim 1/k^2$ and $\sim 1/k^4$) and how they can be measured experimentally by radio-frequency and photo-association spectroscopies. Finally, we evaluate the two contacts at high temperature with a virial expansion. 
\end{abstract}
\maketitle

{\em Introduction}. In the past decade, degenerate Fermi gases close to scattering resonances have attracted both theoretical and experimental attention~\cite{Zwerger2011}. In the unitary Fermi gas close to an $s$-wave resonance, it is understood that thermodynamic properties are universal~\cite{Ho2004}, depending only on a single function, called the ``contact"~\cite{Tan2008,Braaten2008,Zhang2009,Werner2009}. Its manifestations in physical properties have been extensively explored and confirmed in experiments~\cite{Stewart2010,Sagi2012,Hoinka2013}. Extension to arbitrary dimensions has been considered \cite{Valiente2011,Valiente2012}. However, so far contact has only been considered for s-waves, even though p-wave and higher partial wave resonances have been explored experimentally~\cite{Regal2003,Zhang2004,Gunter2005,Schunck2005,Gaebler2007,Fuchs2008,Inada2008,Nakasuji2013,Ticknor2004,Chevy2005,Levinsen2007,Levinsen2008} and theoretically~\cite{Pricoupenko2006,Lasinio2008,Zhang2010,Braaten2012,Nishida2013,Zinner2014,Nishida2014,Gao2014,Peng2014,Gridnev2014,Jiang2015,Ohashi2005,Gurarie2007,Gubbels2007,Inotani2012}.


In this Letter, motivated by the recent radio-frequency (rf) spectroscopic data near a $p$-wave Feshbach resonance in $^{40}$K \cite{Luciuk2015}, we generalize the concept of contact to $p$-waves. In the case of an $s$-wave resonance, a single contact, which depends on the $s$-wave scattering length $a_s$, is sufficient for the characterization of universal properties of the system. For example,  the two-body binding energy is given by $\hbar^2/Ma_s^2$, where $M$ is the mass of the atoms, and the effective range correction is in general small~\cite{Werner2012}. In the case of $p$-wave scattering, however, the phase shift is given by $\cot\delta(k)=-1/vk^3-1/Rk$ for a short-range potential, and the effective range $R$ is of fundamental relevance, in addition to the scattering volume $v$. This can be seen clearly in the binding energy of a shallow $p$-wave bound state $E_b=\hbar^2R/Mv$~\cite{com1}, depending crucially on {\em both} $v$ and $R$. As a result, to capture the universal properties of a spinless Fermi gas around a $p$-wave resonance, it is necessary to introduce two contacts, related to the variation of $v$ and $R$, separately. We show how two adiabatic theorems [see Eqs.~(\ref{eqn:twodensity}) and ({\ref{at})] can be established and how the two contacts relate to the leading ($\sim1/k^2$) and the sub-leading ($\sim1/k^4$) terms of the high-momentum distribution. We also show how the two contacts can be measured spectroscopically. Finally, we use a virial expansion to determine each contact as a function of $T$, $v$, and $R$ at high temperature.


{\em General Formulation}. To start, let us consider the two-body problem, where two identical fermions of mass $M$ interact via a short-range potential $U(r)$ of range $r_0$, tuned close to a $p$-wave resonance. The relative wave function in the $p$-wave channel can be written as $\psi_k({\bf r})\equiv \chi_k(r)Y_{1 m}(\hat{r})/r$, where $m$ labels the projection of angular momentum along $\hat{z}$-direction and $k$ is the relative wave vector. For low-energy $p$-wave scattering, the radial wave function $\chi_k(r)$ can be expanded in powers of $k^2$, $\chi_k(r)\equiv \chi^{(0)}(r)+k^2 \chi^{(1)}(r)+\cdots$. In the asymptotic regime where $1/k\gg r\gg r_0$, we fix the normalization such that the explicit form of $\chi_k(r)$ [and hence $\chi^{(0)}$ and $ \chi^{(1)}$] is  
\be
\chi_k(r)=\left(\frac{1}{r}-\frac{r^2}{3v}\right)+k^2\left(\frac{r}{2}-\frac{r^2}{3R}+\frac{r^4}{30v}\right)+\cdots.
\label{eqn:pwavewf}
\ee
It is important to note that the above asymptotic forms for $\chi^{(0,1)}(r)$ also hold for any shallow $p$-wave bound state in the corresponding asymptotic regime. Once the asymptotes are determined through Eq.~(\ref{eqn:pwavewf}), the short-range form ($r< r_0$) of $\chi^{(0,1)}(r)$ is completely fixed by two-body physics, due to competition between kinetic and potential energy and in particular, independent of the asymptotic wave vector $k$ \cite{Zhang2009}. 

To proceed to the many-body case, we first need to derive two important identities, relating the change of $v$ and $R$ to that of the variation of the potential $U(r)$. Consider two slightly different potentials $U_{\pm}(r)=U(r)\pm \delta U(r)/2$, each with scattering volume $v_{\pm}$ and effective range $R_{\pm}$. The radial Schr\"{o}dinger equation is
\begin{align}
\left(-\frac{\hbar^2}{M}\frac{d^2}{dr^2}+U_{\pm}(r)+\frac{2\hbar^2}{M r^2}\right)\chi_{\pm}(r) &=\frac{\hbar^2 k_{\pm}^2}{M}\chi_{\pm}(r).
\end{align}
The term ${2\hbar^2}/(M r^2)$ gives the $p$-wave centrifugal potential. Following the standard procedure~\cite{Zhang2009,com1}, we find
\begin{align}
\label{eqn:vinv}
\delta v^{-1} &=-\frac{M}{\hbar^2}\int_0^\infty dr \,\delta U(r)|\chi^{(0)}(r)|^2,\\
\delta R^{-1} &=-\frac{2M}{\hbar^2}\int_0^\infty dr \,\delta U(r)\chi^{(0)}(r)\chi^{(1)}(r),
\label{eqn:Rinv}
\end{align}
where $\delta v^{-1}=v_+^{-1}-v_-^{-1}$ and similarly for $\delta{R}^{-1}$. 

Next, consider a spinless Fermi system of total number $N$ confined in volume $\Omega$ with density $n\equiv N/\Omega\equiv k_{\rm F}^3/(6\pi^2)$ where $k_{\rm F}$ is the Fermi wave vector. The two-body density matrix $\rho_2({\bf r}_1,{\bf r}_2)\equiv \langle\psi^\dagger({\bf r}_1)\psi^\dagger({\bf r}_2)\psi({\bf r}_2)\psi({\bf r}_1)\rangle$, where $\psi^\dagger({\bf r})$ creates a fermion at position ${\bf r}$, is Hermitian and can be diagonalized
\be
\rho_2({\bf r}_1,{\bf r}_2)=\sum_\alpha n_\alpha\phi^*_\alpha({\bf r}_1,{\bf r}_2)\phi_\alpha({\bf r}_1,{\bf r}_2).
\ee
The eigenvalues $n_\alpha$ satisfy the condition $\sum_\alpha n_\alpha=N(N-1)$, and the associated pair wave functions $\{\phi_\alpha({\bf r}_1,{\bf r}_2)\}$ form an orthonormal set. In a rotationally invariant system, they can be further written as
\be
\phi_\alpha({\bf r}_1,{\bf r}_2)=\frac{1}{\sqrt{\Omega}r}\exp(i{\bf P}\cdot{\bf R})\varphi_{j\ell}(r)Y_{\ell m}(\hat{r}),
\label{eqn:2bd}
\ee
where ${\bf R}=({\bf r}_1+{\bf r}_2)/2$ is the center of mass and ${\bf r}={\bf r}_1-{\bf r}_2$ is the relative coordinate and $r=|{\bf r}|$. ${\bf P}$ can be regarded as the center-of-mass momentum of a pair and $j,\ell,m$ label the quantum numbers of the relative radial direction, the angular momentum, and its $\hat{z}$ projection respectively. Here the index $\alpha=\{{\bf P},j,\ell,m\}$ is a shorthand for all the quantum numbers that label the pair wave function. In a single-component Fermi gas, $\ell$ must be odd. In the region $r\to 0$, $\varphi_{j\ell}(r)\sim r^{\ell+1}$; the $p$-wave channel has the strongest penetration inside the interaction potential $U(r)$. As a result, we shall concentrate only on the $p$-wave component, since it gives the dominant contribution to the interaction energy of the system.

The pair wave function $\phi_\alpha({\bf r}_1,{\bf r}_2)$, and hence $\varphi_{j\ell}(r)$ is {\em not} an eigenfunction of the two-body Schr\"{o}dinger equation, but can be expanded in terms of the $p$-wave functions (setting $\ell=1$ and neglecting the subscript $\ell$ from $\varphi_{j\ell}$ thereafter)
\be
\varphi_{j}(r)=\int_0^\infty dk a_{jk}\chi_k(r)+a_{j\kappa}\chi_\kappa(r),
\label{eqn:pair}
\ee
where $\{a_{jk},a_{j\kappa}\}$ are the real expansion coefficients, the integration is over all scattering states, and we have also taken into account the possibility of a shallow bound state with radial wave function $\chi_\kappa(r)$ and binding energy $E_b=\hbar\kappa^2/M=\hbar^2 R/(Mv)$, when $v>0$ and $R>0$. Extension to multiple bound states is straightforward. An important consequence of such considerations is that, in the asymptotic region where $r_0\ll r\ll k_{\rm F}^{-1}$, the form of $\chi_k(r)$, and hence $\varphi_{j}(r)$, when expanded in power of $k^2$, are identical to that of $\chi^{(0)}$ and $ \chi^{(1)}$. Furthermore, for $r<r_0$, both are uniquely fixed by the two-body physics. Thus, when evaluating the expectation value of any short-range function such as potential $U(r)$,  $\chi^{(0,1)}$ can be taken out of the integration over $k$.

The interaction energy of the many-body system can be written in terms of $\rho_2$ as $\langle \mathcal{U}\rangle=\frac{1}{2}\int U(|{\bf r}_1-{\bf r}_2|)\rho_2({\bf r}_1,{\bf r}_2)d^3{\bf r}_1d^3{\bf r}_2$. Using the decomposition Eq.~(\ref{eqn:2bd}) and Eq.~(\ref{eqn:pair}), we find
\be
\langle \mathcal{U}\rangle=\frac{1}{2}\sum_m\Big[C_v^{(m)}\int dr U|\chi^{(0)}|^2+C_R^{(m)}\int dr U\chi^{(0)}\chi^{(1)}\Big],
\label{eqn:interaction}
\ee
where we have defined two $p$-{\em wave contacts} $C_{v,R}^{(m)}$ for each $m$,
\begin{align}
C_v^{(m)} &=\sum_{{\bf P},j}n_{{\bf P},j,m}(\int dk a_{jk})^2,\\
C_R^{(m)} &=\frac{1}{2}\sum_{{\bf P},j}n_{{\bf P},j,m}\int dk\int dk' a_{jk}a_{jk'}(k^2+k'^2).
\end{align}
Here the contribution from possible bound states is implicitly included in the integration over $k$. We note that $C_v^{(m)}$ has dimension of length, while $C_R^{(m)}$ has dimension of inverse length. Just as in the $s$-wave case, $C_{v,R}^{(m)}$ encapsulate all the short-range correlations of the many-body system. As a byproduct, the two-body density matrix for $\mathbf r=\mathbf r_1-\mathbf r_2$ in the asymptotic regime $r_0\ll r\ll k_{\rm F}^{-1}$ can be written as 
\be
\rho_2\left(\frac{{\bf r}}{2},-\frac{{\bf r}}{2}\right)=\frac{1}{\Omega}\sum_m |Y_{1m}(\hat r)|^2\left[\frac{C_v^{(m)}}{r^4}+\frac{C_R^{(m)}}{r^2}\right].\label{eqn:twodensity}
\ee

Now suppose that the potential $U(r)$ can be controlled via an auxiliary parameter $\lambda$, such that a small change in $U(r)$ can be written as $(dU/d\lambda)d\lambda$. We can use the Hellmann-Feynman theorem and write $dE/d\lambda=\langle d \mathcal{H}/d\lambda\rangle=\langle d \mathcal{U}/d\lambda\rangle$, where $\mathcal{H}=\mathcal{K}+\mathcal{U}$ is the total many-body Hamiltonian with $\mathcal{K}$ denoting the kinetic energy, independent of $\lambda$. Using Eqs.~(\ref{eqn:vinv},\ref{eqn:Rinv},\ref{eqn:interaction}), we find
\be
\left.\frac{dE}{dv^{-1}}\right|_{R}=-\frac{\hbar^2}{2M}\sum_m C_v^{(m)}, \left.\frac{dE}{dR^{-1}}\right|_{v}=-\frac{\hbar^2}{2M}\sum_m C_R^{(m)}. \label{at}
\ee
In the simplest case of a shallow $p$-wave two-body bound state, the  wave function in the asymptotic region $1/\kappa\gg r\gg r_0$ is given by $\psi_\kappa({\bf r})=\sqrt{R}(1/r^2+\kappa/r)\exp(-\kappa r)Y_{1m}(\hat{r})$ with $\kappa=\sqrt{R/v}$. It is then easy to obtain $\rho_2({\bf r}/2,-{\bf r}/2)=2/\Omega(R/r^4-R^2/vr^2)|Y_{1m}(\hat{r})|^2$ for the two-body bound state~\cite{com1}. One can extract directly that $C_v^{(m)}=2R$ and $C_R^{(m)}=-2R^2/v$, consistent with the adiabatic theorems Eq.~(\ref{at}). The derivation also applies to thermal equilibrium, in which case, one should replace the energy $E$ by the free energy $F$ of the system and keep temperature constant~\cite{Zhang2009}. 

As in the $s$-wave case, a pressure relation and virial theorem can be found. In a uniform system, the universal hypothesis is that the free energy can be written as $F(T/T_{\rm F},k_{\rm F}^3v,k_{\rm F}R)$ close to a $p$-wave resonance, where $T_{\rm F}$ is the Fermi temperature. Using dimensional analysis~\cite{Tan2008, Braaten2008, Zhang2009}, 
\begin{align}
P=\frac{2}{3}\mathcal{E}+\frac{\hbar^2}{2M\Omega v}\sum_mC_v^{(m)}+\frac{\hbar^2}{6M\Omega R}\sum_mC_R^{(m)},
\end{align}
where $\mathcal{E}\equiv E/\Omega$ is the energy density and $E$ is the total energy.  In an external harmonic trap $V({\bf r})=\frac{1}{2}M\omega^2|{\bf r}|^2$,  the free energy can be written as $F(T/T_{\rm F},k_{\rm F}^3v,k_{\rm F}R,\hbar\omega/E_{\rm F})$ near resonance, and we find
\begin{align}
E=2\langle V\rangle-\frac{3\hbar^2}{4Mv}\sum_mC_v^{(m)}-\frac{\hbar^2}{4MR}\sum_mC_R^{(m)},
\end{align}
where $\langle V\rangle$ denote the total potential energy due to the harmonic confinement.

{\em Momentum Distribution}. The correlations encapsulated by the two contacts determine the tail of momentum distribution, which can be measured using time-of-flight imaging~\cite{Stewart2010}. Theoretically,
the momentum distribution can be obtained by Fourier transforming the single-particle density matrix $\rho_1({\bf r},{\bf r}')\equiv N^{-1}\int d^3{\bf r''}\langle\psi^\dagger(\bf r)\psi^\dagger(\bf r'')\psi(\bf r'')\psi(\bf r')\rangle$. While the leading term of momentum distribution is entirely determined by the internal structure of the two-body density matrix in the asymptotic regime [cf.~Eq.~(\ref{eqn:twodensity})], the sub-leading term depends on the distribution of center of mass momentum ${\bf P}$. Assuming that the distribution of ${\bf P}$ is isotropic, namely, $\langle P_x^2\rangle=\langle {\bf P}^2\rangle/3$ {\rm etc.}, we find that, in fact for $k_{\rm F}\ll k\ll 1/r_0$ \cite{com1}
\begin{align} \label{nk}
n_{\bf k}&=\sum_m\left[\frac{16\pi^2 C_v^{(m)}}{\Omega k^2}+\frac{32\pi^2 C_R^{(m)}}{\Omega k^4}\right]|Y_{1m}(\hat{k})|^2\\\nonumber
&-\frac{8\pi^2}{3}\frac{\sum_{\mathbf P,j}n_{\mathbf P,j,m} (\int dq a_{jq})^2 P^2}{\Omega k^4}|Y_{1m}(\hat{k})|^2\\\nonumber
&+\frac{4\pi^2}{3}\frac{\sum_{m,\mathbf P,j}n_{\mathbf P,j,m} (\int dq a_{jq})^2 P^2}{\Omega k^4},
\end{align}
which shows that while $\sum_m C_v^{(m)}$ determines the strength of the leading $1/k^2$~\cite{Inotani2012}, the sub-leading term $1/k^4$ is not determined solely by the contact $\sum_m C_R^{(m)}$. In fact, the angular dependences of the momentum distribution is also not purely $p$-wave, but has an $s$-wave component. We also note that a sub-leading term in momentum distribution relating to the $s$-wave effective range is found in Ref.~\cite{Werner2012}. 

{\em Radio-frequency Spectroscopy}. The rf coupling $H_{\rm rf}=\hbar\Omega_{\rm rf}\int d{\bf r}\psi_e^\dagger({\bf r})\psi({\bf r})$ transfers fermions into an initially empty spin state $|e\rangle$, where $\Omega_{\rm rf}$ is the rf Rabi frequency. For a perturbative $\Omega_{\rm rf}$, the transfer rate can be written as $\Gamma_{\rm rf}(\omega)=(2\pi/\hbar) \sum_{i,f}\rho_i|\langle f|H_{\rm rf}|i\rangle|^2\delta(\hbar\omega+E_i-E_f)$, where $i,j$ label the initial and final states, and $\rho_i$ denotes the initial state distribution. In the region $E_{\rm F}\ll\hbar\omega\ll E_{\rm R}\equiv \hbar^2/MR^2$, one finds~\cite{Punk2007,Schneider2010,Braaten2010}
\be
\Gamma_{\rm rf}(\omega)=\frac{2M\Omega_{\rm rf}^2}\hbar\left[\frac{\sum_mC_v^{(m)}}{(M\omega/\hbar)^{1/2}}+\frac{3\sum_mC_R^{(m)}}{2(M\omega/\hbar)^{3/2}}\right].
\label{eqn:rf}
\ee

{\em Static Structure Factor}. By definition, the static structure factor $S({\bf q})=2\pi\sum_{i,f}\rho_i|\langle f|\int d^3{\bf r}n({\bf r})\exp(-i{\bf q}\cdot{\bf r})|i\rangle|^2$ and can be measured by Bragg spectroscopy~\cite{Vale2013}. Here $n({\bf r})$ is the density operator and other notations are the same as before. $S({\bf q})$ can be obtained directly by Fourier transforming $\rho_2$, Eq.~(\ref{eqn:twodensity}), and diverges {\em linearly} in the limit $q\to\infty$. It is cut off by the short-range potential $U(r)$ and will be limited by $1/r_0$. 

{\em Photo-association Spectroscopy}. Photo-association has been used to measure the fraction of closed channel molecules in two-component Fermi gases~\cite{Partridge2005}, which is related to the $s$-wave contact~\cite{Zhang2009,Werner2009}. In the case of a $p$-wave resonance, if the internal wave function $g_m({\bf r})$ of the relevant excited molecule has a specific projection $m$ along the $\hat z$ direction, namely $g_m({\bf r})\sim Y_{1m}(\hat r)$, the transition rate is given by $\Gamma^{(m)}_{\rm pa}(\omega)=2\pi\hbar\Omega_{\rm pa}^2\sum_{i,f}\rho_i|\langle f|\int d^3{\bf r}d^3{\bf R}g_m^*({\bf r})\phi_m^\dagger({\bf R})\psi({\bf R}+{\bf r}/2)\psi({\bf R}-{\bf r}/2)|i\rangle|^2\delta(\hbar\omega+E_i-E_f)$, with $\Omega_{\rm pa}$ the Rabi frequency and $\phi_m^\dagger({\bf R})$ the molecule creation operator.

Since usually the final molecular state has a finite decay rate $\gamma$, $\delta(\hbar\omega+E_i-E_f)$ in the expression of $\Gamma^{(m)}_{\rm pa}(\omega)$ should be replaced by a Lorentzian $(\hbar\gamma/2)/[(\hbar\omega+E_i-E_f)^2+(\hbar\gamma/2)^2]$. Typically $\gamma\sim 10$ MHz~\cite{Partridge2005}, much larger than the energy scales associated with the spatial motion of the Fermi gas. As a result, when the $\omega$ of the photo-association laser is tuned to resonance, $\hbar\gamma$ dominates over typical values of $\hbar\omega+E_i-E_f$, and we can approximate the Lorentzian by $2/\hbar\gamma$,
\be
\Gamma^{(m)}_{\rm pa}=\frac{4\pi}{\gamma}C_v^{(m)}\Omega_{\rm pa}^2\left|\int d^3{\bf r}g_m^*({\bf r})Y_{1m}(\hat{r})\frac{\chi^{(0)}(r)}{r} \right|^2. \label{pa}
\ee
The Franck-Condon factor can be computed once $g_m({\bf r})$ is known. What is important here is that it depends only on two-body physics, so the many-body dependence is encapsulated in $C_v^{(m)}$. The contribution from $C_R^{(m)}$ is smaller by a factor $(k_{\rm F}r_0)^2$ if the excited molecular state is of extension $r_0$. In the case when the photo-association process does not distinguish between final molecular states of different $m$, the total transition rate will be the sum of the individual $\Gamma^{(m)}_{\rm pa}$. 

{\em Virial Expansion for $p$-wave Contacts}. At high temperatures, the effects of the interaction can be taken into account by the second virial expansion \cite{HM2004}. The change of the free energy of the spinless fermions $\delta F\equiv F-F_0$ is given by $\delta F/k_{\rm B}T=-2\sqrt{2}Nn\lambda^3b_2$, where $F_0$ is the free energy without interactions and $\lambda\equiv h/\sqrt{2\pi Mk_{\rm B}T}$. 
The second virial coefficient is given by
\be
b_2=3\left[\int_0^\infty \frac{dk}{\pi}\frac{d\delta(k)}{dk}e^{-\lambda^2k^2/2\pi}+\theta(v)e^{E_b/k_{\rm B}T}\right].\label{b2}
\ee

Let $C_{v,R}\equiv \sum_mC_{v,R}^{(m)}$, then by the adiabatic theorems,
\begin{align}
\frac{C_v}{N}=8\sqrt{2}\pi n\lambda\frac{\partial b_2}{\partial v^{-1}},~\frac{C_R}{N}=8\sqrt{2}\pi n\lambda\frac{\partial b_2}{\partial R^{-1}}.
\end{align}
When $v^{-1}=0$, $\partial b_2/\partial v^{-1}=(3/\sqrt{2\pi})\lambda R^2 h_v(\lambda/(R\sqrt{2\pi}))$ with
\begin{align}
h_v(\eta)
=\eta+\eta^2\int_0^\infty dx \frac{(1-e^{-x^2})(\eta^2+3x^2)}{\pi(\eta^2 x+x^3)^2},\label{hv}
\end{align}
and $\partial b_2/\partial R^{-1}=(3/\sqrt{2\pi})\lambda h_R(\lambda/(R\sqrt{2\pi}))$ with
\begin{align}
h_R(\eta)= \frac1{\sqrt{\pi}}-\eta e^{\eta^2} {\rm Erfc}(\eta).\label{hr}
\end{align}
Figure \ref{ht} shows the dependences of $C_v$ and $C_R$ as a function of $E_b/E_{\rm F}$ for $T/T_{\rm F}=2$ and $k_{\rm F}R=1/25$, appropriate for the case of $^{40}$K with a typical density of $2\times 10^{19}$ m$^{-3}$ at the $p$-wave resonances near $B=198.5$G~\cite{Regal2003}. Here we note that while $C_v$ decreases monotonically as $-E_b/E_{\rm F}$ increases, $C_R$ shows non-monotonic behavior and reaches a maximum when $-E_b/E_{\rm F}\sim 2$, where it is comparable to $C_v$, if non-dimensionalized by $k_{\rm F}$ (see Fig.~\ref{ht}). The temperature dependence of the contacts $C_v$ and $C_R$ at $v^{-1}=0$ is shown in the inset, for which $C_R$ is much smaller than $C_v$; the magnitude of both grows with increasing $R$.  In the temperature regime $T_{\rm F}\lesssim T\ll \hbar^2/2MR^2$, Eqs.~(\ref{hv}) and (\ref{hr}) give $C_v\approx 12\sqrt{2}Nn\lambda^3R\sim T^{-3/2}$ and $C_R\approx24\pi Nn R^2(1-3R^2Mk_{\rm B}T/2\hbar^2)$, which should be contrasted with $T^{-1}$-dependence for $s$-wave contact~\cite{Yu2009}.

Away from $p$-wave resonances where the scattering volume $v$ is small, Eq.~(\ref{b2}) gives $C_v=72\pi^2Nnv^2/\lambda^2$ and $C_R=360\pi^3 Nnv^2/\lambda^4$ when contribution from the deeply bound state is excluded. In this limit, the scaling $C_v\sim v^2$ and $C_R\sim v^2$ is also expected from perturbation calculations for small $v$~\cite{Pricoupenko2006}, which indicates the irrelevance of $C_v$ and $C_R$ in Fermi gases close to an $s$-wave resonance.

\begin{figure}
\begin{center}
\includegraphics[width=\columnwidth]{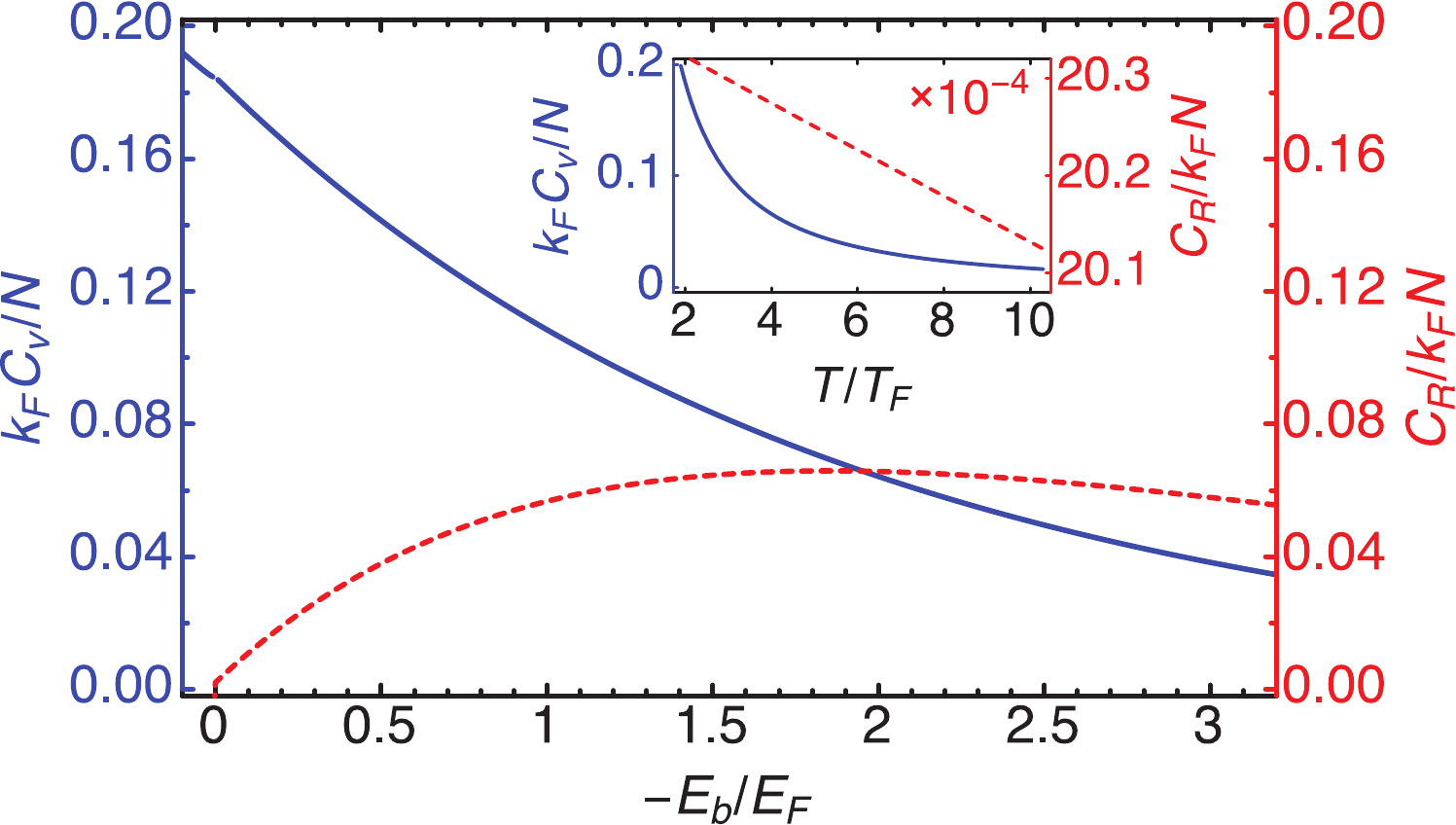}
\end{center}
\caption{(Color online.) Dependence of the contacts $C_v$ (the solid line) and $C_R$ (the dashed line) on $E_b/E_{\rm F}$ for $T/T_{\rm F}=2$ and their temperature dependences (inset) at a $p$-wave resonance from the virial expansion. We have taken $k_{\rm F}R=1/25$.}
\label{ht}
\end{figure}

{\em Discussion}. 
Our derivation of the $p$-wave contacts is based on the single-channel model which does not take into account explicitly the presence of closed-channel molecules, as in the case of a Feshbach resonance. The same results shall be obtained for a two-channel model, provided that the closed-channel molecule is small (comparable to $r_0$), which is typically the case. This is because all the arguments so far depend only on the properties of the two-body wave function or two-body density matrix in the asymptotic regime, which, in our derivation, depend only on the scattering volume $v$ and effective range $R$, irrespective of whether they arise from a shape resonance or a Feshbach resonance. For actual atomic systems, the van der Waals potential modifies the $p$-wave scattering phase shift by introducing a term $\alpha/k^2$ in the effective range expansion~\cite{Bo1998, Zhang2010}. However, close to a Feshbach resonance, it was shown that $\alpha\sim1/v^2$, whose effects are thus negligible~\cite{Zhang2010,Bo2011}. As a result,  we expect that our main results Eqs.~(\ref{eqn:twodensity}) to (\ref{pa}) to remain true close to a $p$-wave Feshbach resonance.   

 
Resonances for different $|m|$ can be split due to magnetic dipole-dipole couplings~\cite{Ticknor2004}. For $^{40}$K, the $m=0$ and $m=\pm 1$ resonances  around $B=198.5$ G are split by about $0.5$ G~\cite{Ticknor2004}. To take this into account, we can introduce phase shifts for different $m$, $\cot\delta^{(m)}=-1/v^{(m)}k^3-1/R^{(m)}k$. Likewise, we can establish the relation $d E/d(1/v^{(m)})=-\hbar^2 C_v^{(m)}/2M$ and $d E/d(1/R^{(m)})=-\hbar^2 C_R^{(m)}/2M$ while Eqs.~(\ref{eqn:twodensity}) and (\ref{nk}) to (\ref{pa}) stay intact.


\emph{Acknowledgements.} We are grateful to Shina Tan, Chris Luciuk and Stefan Trotzky for useful discussions; to Pengfei Zhang for correcting an error in Eq.~(\ref{eqn:rf}); and to W. Zwerger for drawing our attention to Ref.~\cite{Werner2012}. ZY is supported by NSFC under Grant No. 11474179 and 11204152, and the Tsinghua University Initiative Scientific Research Program. JT is supported by AFOSR, ARO, and NSERC. SZ is supported by GRF HKU 17306414 and CRF HKUST3/CRF/13G, and the Croucher Foundation under the Croucher Innovation Award. 

{\em Note added}. During the final preparation of this manuscript, closely related work by Yoshida and Ueda appeared~\cite{Yoshida2015}, in which they discuss one of the contacts, $C_v$, using a two-channel model.


\begin{thebibliography}{100}
\bibitem{Zwerger2011}
W. Zwerger, ed., {\it The BCS-BEC Crossover and the Unitary Fermi Gas}, (Springer-Verlag 2011).

\bibitem{Ho2004}
Tin-Lun Ho, Phys. Rev. Lett. {\bf 92} 090402 (2004).

\bibitem{Tan2008}
Shina Tan, Ann. Phys. N.Y. {\bf 323}, 2952 (2008).

\bibitem{Braaten2008}
E. Braaten and L. Platter, Phys. Rev. Lett. {\bf 100}, 205301 (2008).

\bibitem{Zhang2009}
S. Zhang and A. J. Leggett, Phys. Rev. A {\bf 79}, 023601 (2009).

\bibitem{Werner2009}
F. Werner, L. Tarruell, Y. Castin, Euro. Phys. J. B {\bf 68}, 410 (2009). 

\bibitem{Stewart2010}
J. T. Stewart, J. P. Gaebler, T. E. Drake, and  D. S. Jin, Phys. Rev. Lett.  {\bf 104}, 235301 (2010).

\bibitem{Sagi2012}
Y. Sagi, T. E. Drake, R. Paudel, and D. S. Jin, Phys. Rev. Lett. {\bf 109}, 220402, (2012).

\bibitem{Hoinka2013}
S. Hoinka, M. Lingham, K. Fenech, H. Hu, C. J. Vale, J. E. Drut, and S. Gandolfi, Phys. Rev. Lett. {\bf 110}, 055305 (2013).

\bibitem{Valiente2011}M. Valiente, N.T. Zinner and K. M{\o}lmer, Phys. Rev. {\bf84}, 063626 (2011).

\bibitem{Valiente2012}M. Valiente, N.T. Zinner and K. M{\o}lmer, Phys. Rev. A {\bf86}, 043616 (2012).
\bibitem{Regal2003}
C. A. Regal, C. Ticknor, J. L. Bohn, and D. S. Jin, Phys. Rev. Lett. {\bf 90}, 053201 (2003).

\bibitem{Zhang2004}
J. Zhang, E. G. M. van Kempen, T. Bourdel, L. Khaykovich, J. Cubizolles, F. Chevy, M. Teichmann, L. Tarruell, S. J. J. M. F. Kokkelmans, and C. Salomon, Phys. Rev. A {\bf 70}, 030702(R) (2004).

\bibitem{Gunter2005}
Kenneth G\"{u}nter, Thilo St\"{o}ferle, Henning Moritz, Michael K\"{o}hl, and Tilman Esslinger, Phys. Rev. Lett. {\bf 95}, 230401 (2005).

\bibitem{Schunck2005}
C. H. Schunck, M. W. Zwierlein, C. A. Stan, S. M. F. Raupach, W. Ketterle, A. Simoni, E. Tiesinga, C. J. Williams, and P. S. Julienne, Phys. Rev. A {\bf 71}, 045601 (2005).

\bibitem{Gaebler2007}
J. P. Gaebler, J. T. Stewart, J. L. Bohn, and D. S. Jin, Phys. Rev. Lett. {\bf 98}, 200403 (2007).

\bibitem{Fuchs2008}
J. Fuchs, C. Ticknor, P. Dyke, G. Veeravalli, E. Kuhnle, W. Rowlands, P. Hannaford, and C. J. Vale, Phys. Rev. A {\bf 77}, 053616 (2008).

\bibitem{Inada2008} 
Y. Inada, M. Horikoshi, S. Nakajima, M. Kuwata-Gonokami, M. Ueda, and T. Mukaiyama, Phys. Rev. Lett. {\bf 101}, 100401 (2008).

\bibitem{Nakasuji2013}
Takuya Nakasuji, Jun Yoshida, and Takashi Mukaiyama, Phys. Rev. A {\bf 88} 012710 (2013).


\bibitem{Ticknor2004}
C. Ticknor, C. A. Regal, D. S. Jin, and J. L. Bohn, Phys. Rev. A {\bf 69}, 042712 (2004).

\bibitem{Chevy2005}
F. Chevy, E.G.M. van Kempen, T. Bourdel, J. Zhang, L. Khaykovich, M. Teichmann, L. Tarruell, S.J.J.M.F. Kokkelmans, C. Salomon, Phys. Rev. A {\bf 71}, 062710 (2005).

\bibitem{Levinsen2007}
J. Levinsen, N. R. Cooper,  and V. Gurarie, Phys. Rev. Lett. {\bf 99}, 210402 (2007).

\bibitem{Levinsen2008}
J. Levinsen, N. R. Cooper,  and V. Gurarie, Phys. Rev. A {\bf 78}, 063616 (2008).


\bibitem{Pricoupenko2006}
 L. Pricoupenko, Phys. Rev. Lett. {\bf 96}, 050401 (2006).

\bibitem{Lasinio2008}
M. Jona-Lasinio, L. Pricoupenko, and Y. Castin, Phys. Rev. A {\bf 77}, 043611 (2008).

\bibitem{Zhang2010}
P. Zhang, P. Naidon, and M. Ueda, Phys. Rev. A {\bf 82}, 062712 (2010).

\bibitem{Braaten2012}
E. Braaten, P. Hagen, H.-W. Hammer, and L. Platter, Phys. Rev. A {\bf 86}, 012711 (2012).

\bibitem{Nishida2013}
Y. Nishida, S. Moroz, and D. T. Son, Phys. Rev. Lett. {\bf 110}, 235301 (2013).

\bibitem{Zinner2014}
A. G. Volosniev, D. V. Fedorov, A. S. Jensen, N. T. Zinner,  J. Phys. B {\bf 47}, 185302 (2014).

\bibitem{Nishida2014}
S. Moroz and Y. Nishida, Phys. Rev. A {\bf 90}, 063631 (2014).

\bibitem{Gao2014}
C. Gao, J. Wang and Z. Yu, arXiv:1412.3566.

\bibitem{Peng2014}       
S. G. Peng, S. Tan and K. Jiang, Phys. Rev. Lett. {\bf 112}, 250401 (2014).

\bibitem{Gridnev2014}
D.K.~Gridnev, J. Phys. A: Math. Theor. {\bf 47} 505204 (2014)

\bibitem{Jiang2015}
T. Y. Gao, S. G. Peng, and K. Jiang, Phys. Rev. A {\bf91}, 043622 (2015).

\bibitem{Ohashi2005}
Y. Ohashi, Phys. Rev. Lett. {\bf 94}, 050403 (2005).

\bibitem{Gurarie2007}
V. Gurarie, L. Radzihovsky, Annals of Physics {\bf 322}, 2 (2007).

\bibitem{Gubbels2007}
K. B. Gubbels  and H. T. C. Stoof, Phys. Rev. Lett. {\bf 99}, 190406 (2007).

\bibitem{Inotani2012}
D. Inotani, R. Watanabe, M. Sigrist, and Y. Ohashi, Phys. Rev. A {\bf 85}, 053628 (2012).

\bibitem{Luciuk2015} 
C. Luciuk, S. Trotzky, S. Smale, Zhenhua Yu, Shizhong Zhang and J. H. Thywissen, arXiv: 1505.08151 (2015). 


\bibitem{Werner2012}
F. Werner and Y. Castin, Phys. Rev. A, {\bf 86}, 013626 (2012).

\bibitem{com1} 
See supplementary materials for derivations of the shallow $p$-wave two-body bound states, derivation of equations (3) and (4), and the tail of momentum distribution.


\bibitem{Punk2007}
M. Punk and W. Zwerger, Phys. Rev. Lett. {\bf 99}, 170404 (2007).

\bibitem{Schneider2010}
W. Schneider and M. Randeria, Phys. Rev. A {\bf 81}, 021601 (2010).

\bibitem{Braaten2010}
E. Braaten, D. Kang, L. Platter, Phys. Rev. Lett. {\bf 104}, 223004 (2010)

\bibitem{Vale2013}
S. Hoinka, M. Lingham, K. Fenech, H. Hu, C.J. Vale, J.E. Drut, and S. Gandolfi, Phys. Rev. Lett. \textbf{110}, 055305 (2013).

\bibitem{Partridge2005}
G. B. Partridge, K. E. Strecker, R. I. Kamar, M. W. Jack, and R. G. Hulet, Phys. Rev. Lett. {\bf 95}, 020404 (2005).

\bibitem{HM2004}
T-L Ho and E.J. Mueller, Phys. Rev. Lett. \textbf{92}, 160404 (2004).

\bibitem{Yu2009}
Z. Yu, G.M. Bruun, and G. Baym, Phys. Rev. A \textbf{80}, 023615 (2009).


\bibitem{Bo1998}
B. Gao, Phys. Rev. A {\bf 58}, 4222 (1998).

\bibitem{Bo2011}
B. Gao, Phys. Rev. A {\bf 84}, 022706 (2011).

\bibitem{Yoshida2015} 
Shuhei M. Yoshida and Masahito Ueda, Phys. Rev. Lett. {\bf 115}, 135303 (2015)


\end{thebibliography}
\end{document}